\documentclass[aps,prb,twocolumn,showpacs,groupedaddress,amsmath,,floatfix,amssymb]{revtex4-2}
\usepackage{epsfig}
\usepackage{dcolumn}
\usepackage{graphicx}
\usepackage{amsmath}
\usepackage{amsfonts}
\usepackage{amssymb}
\usepackage{color}
\usepackage{txfonts}
\usepackage{hyperref}
\usepackage{adjustbox}
\usepackage{ulem}
\usepackage{framed}
\usepackage{wrapfig}
\usepackage{multirow}
\usepackage{xcolor}

\newcommand{\be}{\begin{equation}}
\newcommand{\ee}{\end{equation}}
\newcommand{\bea}{\begin{eqnarray}}
\newcommand{\eea}{\end{eqnarray}}

\def\k{{ {\mathbf k} }}

\def\q{{ {\mathbf q} }}
\def\Q{{ {\mathbf Q} }}
\def \Im{{ \mbox {Im} }}
\def  \Re{{ \mbox {Re} }}

\begin{document}

\title{Phonon spectral function of Holstein polaron: Investigation of many-body effects with self-energy and vertex correction}

\author{Lawrence Rai and Sudhakar Pandey} 
\affiliation{Department of Physics, North-Eastern Hill University, Shillong - 793022, Meghalaya, India}

\begin{abstract}
 We study the impact of the many-body effects on the phonon spectral function of Holstein polaron in one-dimension in the antiadiabatic regime by incorporating  the contributions from the  electron self-energy and vertex corrections within a weak-coupling approach that respects the charge-conserving Ward identity. 
 We find that while the polaronic spectral weight is suppressed  due to contribution from the electron self-energy, on the other hand, the same is enhanced due to  contribution from the vertex corrections. While strength of both the contributions increases with increasing the wave vector ($\q$) of phonons, they nearly cancel each other for the small-$\q$ modes  so that the polaronic spectral weight is weakly affected  due to  the many-body effects. For  the large-$\q$ modes  near the zone boundary, the  net many-body correction is dominated by the contribution of the electron self-energy which increases faster in comparison to that of the vertex corrections  with increasing the  wave vector thereby resulting in a significant suppression of the polaronic spectral weight. We find that while the weak-coupling perturbative approach provides a reliable estimation of the impact of the many-body effects deep inside the antiadibatic regime, the renormalization of quasiparticle spectrum must be taken into account for an accurate estimation when the phonon energy approaches the electronic bandwidth. 

 \end{abstract}

\maketitle

\section {Introduction}
 A polaron represents a quasiparticle consisting of an electron surrounded by the cloud of lattice vibrations (phonons). The formation of polarons  is one of the most important manifestations of  the electron-phonon coupling (EPC) in materials \cite{Polaron-Book-2010}. Polarons are speculated to play crucial roles in governing several novel phenomena in materials which include, for example, superconductivity, charge transport, colossal magnetoresistance, thermoelectricity,  etc.\cite{ Franchini-NM-2021}.  Experimental evidence for polaron hallmarks  have also been reported using a variety of techniques  in case of several materials,  such  as cuprates, manganites, transition metal oxides, diluted magnetic semiconductors, etc. \cite{ Franchini-NM-2021}.

Theoretically, one of the major issues has been to understand the impact of the many-body effects arising due to electron-phonon coupling on the various polaronic properties in materials \cite{Polaron-Book-2010,Franchini-NM-2021,Giustino-PRB-2022}.  Over the years the Holstein model\cite{Holstein-1959-I,Holstein-1959-II}  has emerged as one of the most attractive model  for investigating the many-body effects on polaronic properties owing to its simplicity where both the phonon spectrum as well as the strength of electron-phonon coupling are approximated to be momentum-independent.   Among these  the studies of the spectral function of polaron have been at the forefront\cite{Marsiglio-PhysLettA-1993, Alexandrov-PRB-1994, Ciuchi-PRB-1997, White-PRB-1999, Hohenadler-PRB-2003, Osor-PRB-2004, Perroni-PRB-2005, Loos-JPCM-2006, Berciu-PRB-2006, Berciu-PRB-2007, Trugman-PRB-2010, Berciu-PRB-2019, Veljko-PRB-2022, Tanaskovic-PRL-2022} which provide valuable insights about the various aspects of the  many-body effects in materials that can also be probed experimentally, for example,  using the angle-resolved photoemission spectroscopy (ARPES)\cite{Mingu-Nat-Mater-2018}. Most of such investigations done so far have been focused on  the  electronic spectral function of the single polaron where many-body effects are incorporated by taking into account the renormalization of electron Green's function using a variety of approaches, such as the momentum-average approximation\cite{Berciu-PRB-2006}, dynamical mean-field theory \cite{Ciuchi-PRB-1997}, density-matrix renormalization group\cite{White-PRB-1999}, etc.  On the other hand,  a detail study of the phonon spectral function  of polaron has been done scarcely in the literature which also contains useful information about several characteristic features of the polarons \cite{Loos-JPCM-2006,Trugman-PRB-2010}.  For example, while the bare phonon spectral function in the Holstein model is characterized by a single peak at frequency $\omega=\Omega_0$, where $\Omega_0$ represents the frequency  of the dispersionless 
optical phonons, the electron-phonon coupling induced formation of  polaron results in emergence of the finite spectral weight at frequencies both below and above  $\Omega_0$,  details of which depend upon the various model parameters, such as the strength of electron-phonon coupling, phonon frequency $\Omega_0$, electronic bandwidth, etc.\cite{Loos-JPCM-2006,Trugman-PRB-2010}

In the earlier investigations the issue of the impact of the many-body effects on the phonon spectral function of polaron has only been partially addressed 
 wherein the contribution of the electron self-energy has been taken into account while the vertex corrections were ignored \cite{Loos-JPCM-2006}. In general,  the vertex corrections are frequently ignored in the investigations of the many-body effects in electron-phonon coupled systems owing to the  Migdal's theorem\cite{Migdal-JETP-1958}. However, the universal applicability of the Migdal's theorem has been a matter of intense debate\cite{Allen-Solid-State-1983,Gunnarsson-PRB-1994,Grimaldi-PRB-1995, Grimaldi-PRB-1995II, Miller-PRB-1998,Hague-JPCM-2003,Hague-JPCM-2005,Gunnarsson-PRB-2011, Kivelson-PRB-2018, Aperis-PRB-2020, Aperis-PRB-2021} as this theorem was originally proposed  for the three-dimensional systems assuming that  the  effective electron-phonon coupling   $(\lambda)$ times the adiabaticity ($\Omega/\epsilon_F$) is very small,  i.e.,   $\lambda \Omega/\epsilon_F << 1$, where adiabaticity refers to  the ratio of the characteristic phonon energy  $(\Omega)$  and the Fermi energy $(\epsilon_F)$. For  example,  based on the phase space arguments the violation of  the Migdal's theorem is  speculated  in general for the systems  in  one-  and two-dimensions\cite{Allen-Solid-State-1983} where  the profound impacts of the vertex corrections have indeed been demonstrated in several studies\cite{Aperis-PRB-2020, Hague-JPCM-2005, Hague-JPCM-2003, Gunnarsson-PRB-1994}. 
These include, for example, the appearance of a rich satellite structure in the exact one-electron spectra of a one-dimensional system\cite{Gunnarsson-PRB-1994}  and significant change in both the  size of the superconducting gap and the transition temperature in case  of a two-dimensional system \cite{Aperis-PRB-2020} when the contributions of the vertex corrections are taken into account.

It must be emphasized that the role of vertex corrections becomes of special importance in context of the phonon spectrum where they cancel exactly the contribution of  the electron self-energy for the $\q=0$ mode  as a consequence of the charge-conserving Ward identity, 
as demonstrated  explicitly in a recent study\cite{SP-PRB-2024}.  Furthermore,  the exact cancellation occurs irrespective of the system's microscopic details, such as adiabaticity, dimensionality, strength of electron-phonon coupling, etc., which has important implications on the impact of the many-body effects on the phonon spectrum.  For example, the many-body effects can naively be expected to have weak impact in the long-wavelength limit $(\q \rightarrow 0)$ due to the continuous evolution of the opposite contributions from the vertex corrections and electron self-energy as we go away
 from the $\q = 0$ mode. Also, the vanishing of the phonon
 self-energy for the $\q = 0$ mode due to the exact cancellation is a fundamental requirement for the  acoustic phonons owing to the translational symmetry of crystals in order to preserve their characteristic gapless spectrum associated with the long-wavelength excitations with the $\q=0$ mode being massless. We note that  in absence of the vertex corrections the finite contribution from the electron self-energy will result in a spurious finite phonon self-energy even for the $\q = 0$ mode, as indeed reported in earlier investigations \cite{Loos-JPCM-2006}, which thereby not only provides an overestimation of the many-body effects but also destroys the characteristic gapless spectrum in the long-wavelength regime.

In this paper we  study the impact of the many-body effects  on the electron-phonon coupling driven  polaronic  features in the phonon spectral function by taking into account the momentum- and frequency-dependent evolution of the contributions from the electron self-energy and vertex corrections.  
We focus  our study  in the antiadiabtic regime where the phonon energy exceeds  the electronic energy scale characterized by the electronic bandwidth.  As the  antiadibatic regime also invalidates the application of the Migdal's theorem, therefore, the vertex corrections can naturally be expected to make important contributions. Our study will also be relevant for the systems in which the adiabatic Born-Oppenheimer approximation, where the phonon energy scale is assumed to be much smaller than its electronic counterpart, breaks down. In this context, the low-dimensional materials, such as the layered graphene and transition metal dichalcogenides,  have drawn special attention wherein the phonon spectrum is believed to be dominated by the nonadiabatic effects\cite{Mauri-NatMat-2007,Dresselhaus-PRL-2012,Zhu-PRL-2025, Wang-PRB-2024}.
For example, the nonadiabatic effects in case of the monolayer and bilayer  graphene are speculated to cause deviation from the characteristic parabolic dispersion for the longitudinal optical phonon around  the center of Brillouin zone, as observed recently using the high-resolution electron energy loss spectroscopy\cite{Zhu-PRL-2025}.  Similarly,
in case of the layered compound WS$_2$, the strong renormalization of the zone-edge acoustic phonons in comparison to their zone-center counterparts, as reported using the Raman spectroscopy, is also believed to have its origin in the non-adiabatic effects\cite{Wang-PRB-2024}. 

This paper is organized as follows.  In the next section, we first briefly review  the Ward identity preserving approach for studying the many-body effects on the phonon spectrum using the Holstein model for the general cases and then we extend our analysis  for the specific case of the single polaron in a one-dimensional system. In Sec. III, we analyze the impact of many-body effects on the low energy polaronic features in the phonon spectral function.  In Sec. IV we extend our analysis by taking into account the renormalization of the quasiparticle spectrum in order to make a better estimation of the many-body effects. Finally, conclusions are presented in Sec. V.     

\section{Model}
 We consider the coupling between electrons in a single band  with a branch of the dispersionless optical phonons  characterized by frequency  $\Omega_0$ within the Holstein model 

\bea
\hat{H}&=&\sum_\k\epsilon_\k\;c_\k^{\dagger}c_\k+
\Omega_0 \sum_\q b^\dagger_\q b_\q
\nonumber \\& + &
 \frac{g}{\sqrt{ N}} \sum_{\k,\q}c_{\k+\q}^\dagger c_\k\;(b^{\dagger}_{-\q}+b_\q)
\label{Ham}\;.
\eea

\noindent  Here  $c_\k^{\dagger}$ ($c_\k$) and $b^{\dagger}_\q$ ($b_\q$) denote the creation (annihilation) operators for the electron and phonon, respectively,  $\epsilon_\k$ denotes  the bare electronic  dispersion, $g$ denotes the strength of electron-phonon scattering, and $N$ is the total number of lattice sites so  that $ N\rightarrow \infty$ in the thermodynamic limit. A vertex pair can be associated with the factor $g^2/N$ which can be utilized as an expansion parameter for the diagrammatic expansion\cite{Aperis-PRB-2020}.

\subsection{Ward identity preserving approach}
In order to study the impact of the  many-body effects on the phonon spectrum we  consider the phonon self-energy $\Pi_\q(\omega)$   by taking into account the contributions of both the electron self-energy as well as the vertex corrections as follows.
\bea
\Pi_\q(\omega) &=&- i \frac{g^2}{N}  
\sum_\k \int \frac{d E}{2\pi} G_\k(E) G_{\k+\q}(E+\omega)
\nonumber \\
& \times &
\Gamma_{\k,\k+\q}(E,E+\omega) \;,
\label{Eq-phonon-SE}
\eea

\noindent Here $G$ and $\Gamma$ represent the exact electron Green's function and the full electron-phonon vertex, respectively.  The dynamical phonon self-energy for the $\q=0$ mode vanishes identically, i.e., $\Pi_0(\omega)=0$,  as required by  the charge-conserving Ward identity\cite{SP-PRB-2024, Mahan-Book,Schrieffer-PR-1963,Chubukov-PRB-2005} which provides a specific relation between $\Gamma$  and $\Sigma$ for the $\q=0$ mode,  viz.  $\Gamma_{\k,\k}(E,E+\omega)= 1+[\Sigma_\k(E)-\Sigma_\k (E+\omega)]/\omega$.

As described in Ref. \cite {SP-PRB-2024}, using  the Dyson equation for the electron Green's function
$G_\k(E)=G^0_\k(E) + G^0_\k(E) \Sigma_\k (E)G_\k(E)$, where  $G^0_\k(E)$ represents the bare electron Green's function, and expanding the electron self-energy  $ \Sigma=\Sigma^{(1)}+\Sigma^{(2)}+\Sigma^{(3)}..$, and vertex corrections $\Delta \Gamma (= \Gamma-1)=\Gamma^{(1)}+\Gamma^{(2)}+\Gamma^{(3)} ..$,  in powers of $g^2/N$,  the phonon self-energy can also be expanded in powers of ($g^2/N$) as follows.
\bea
\Pi_\q(\omega)=\Pi^{(1)}_\q(\omega) + \Pi^{(2)}_\q(\omega)+\Pi^{(3)}_\q(\omega)+ ...  \;.
\label{Eq-Phonon-SE-expansion}
\eea
\noindent

While the leading-order contribution  $\Pi^{(1)}$  represents the interaction between bare electrons and  phonons, the many-body effects involving the contributions from the electron self-energy and vertex  corrections are incorporated  by the higher order terms in Eq. (\ref{Eq-Phonon-SE-expansion}).  For simplicity, we analyze the impact of the many-body effects by focusing only on the subleading-order phonon self-energy $\Pi^{(2)}_\q(\omega)$,  which is a  good  approximation in case of the weakly coupled systems.  These two contributions can be expressed as follows.

\begin{widetext}

\begin{figure}
\vspace*{0mm}
\begin{center}
\hspace*{-5mm}
\resizebox{140mm}{110mm}{\psfig{figure=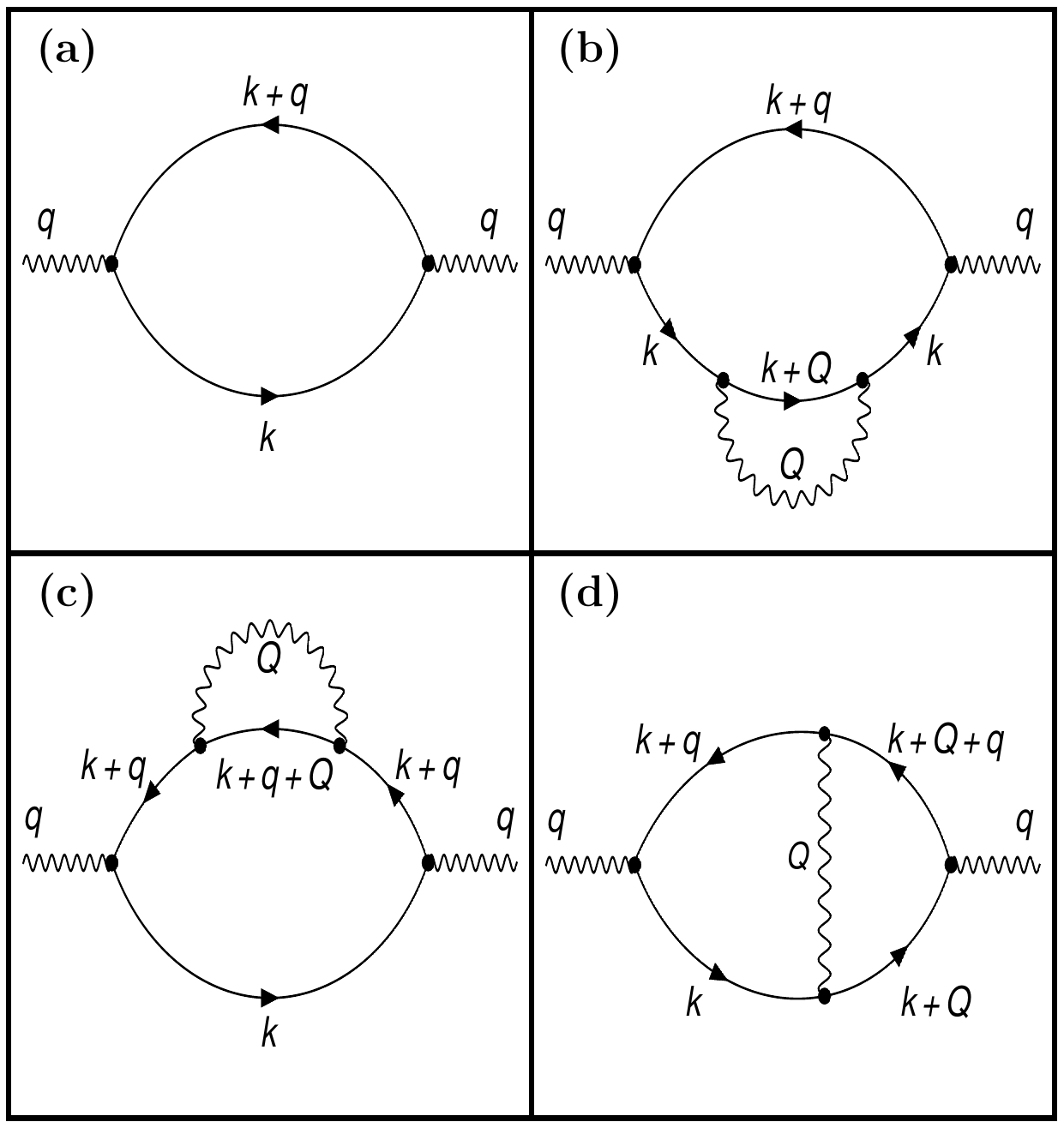}}
\end{center}
\vspace*{-15mm}
\caption{Diagrammatic representations of the  leading-order (a) and subleading-order (b,c,d) phonon self-energies.
Here the single lines with arrow and wavy lines represent the bare electron Green's  function  $G^0$ and the bare  phonon Green's function $D^0$, respectively.  While the leading-order phonon self-energy   $\Pi^{(1)}_\q(\omega)$ (a) represents the coupling between bare electrons  and  phonons, the different components of the subleading-order phonon self-energy involve the contributions from both the electron self-energy, as denoted by $\Pi^{(2a)}_\q(\omega)$ (b) and $\Pi^{(2b)}_\q(\omega)$ (c), as well as the vertex corrections, as denoted by $\Pi^{(2c)}$ (d).  Here we follow the implicit notations $\q=(\q,\omega), \k=(\k,E), \Q=(\Q,\Omega)$.}
\label{Fig-Feynman}
\end{figure}

\bea
\Pi^{(1)}_\q(\omega)& = &-i \frac{g^2}{N} \sum_k \int \frac{d E}{2\pi} G^0_\k(E)G^0_{\k+\q}(E+\omega) \; .
\label{Eq-Pi1}
\eea

\bea
\Pi^{(2)}_\q(\omega)& = &-i \frac{g^2}{N} \sum_k \int \frac{d E}{2\pi} \left \{ [G^0_\k(E)]^2 \Sigma^{(1)}_\k(E)  G^0_{\k+\q}(E+\omega) 
 +    G^0_\k(E) [G^0_{\k+\q}(E+\omega)]^2 \Sigma^{(1)}_{\k+\q}(E+\omega) 
\right. \nonumber \\ 
& + & \left. 
G^0_\k(E)G^0_{\k+\q}(E+\omega) \Gamma^{(1)}_{\k,\k+\q}(E,E+\omega)  \right \}
 \nonumber \\
& =  & \Pi^{(2a)} + \Pi^{(2b)} + \Pi^{(2c)} \; .
\label{Eq-Pi2}
\eea
\noindent 
\end{widetext}

\noindent Here $\Pi^{(2a)}$ and $\Pi^{(2b)}$ represent the impact of the leading-order electron self-energy  $\Sigma^{(1)}$, 
and $\Pi^{(2c)}$ represents the impact of the leading-order vertex corrections  $\Gamma^{(1)}$. The diagrammatic representations for  $\Pi^{(1)}$, $\Pi^{(2a)}$, $\Pi^{(2b)}$, and $\Pi^{(2c)}$  are shown in Fig. (\ref{Fig-Feynman}). Here  $\Sigma^{(1)}$ and  $\Gamma^{(1)}$ can be expressed as, 

\bea
\Sigma^{(1)}_\k(E) &=& i \frac{g^2}{N}
\sum_\Q 
\int \frac {d\Omega}{2\pi}  G^0_{\k+\Q}(E+\Omega)
D^0_\Q(\Omega)
\;  , 
\label{Eq-electron-SE}
\eea

\bea
\Gamma^{(1)}_{\k,\k+\q}(E,E+\omega) 
&=& i \frac{g^2}{N}
\sum_\Q \int \frac{d \Omega}{2\pi} G^0_{\k+\Q}(E+\Omega) 
\nonumber \\
& \times & 
G^0_{\k+\Q+\q}(E+\Omega+\omega) 
D^0_\Q(\Omega)
\;  , 
\eea

\noindent where $D^0_\Q(\Omega)$ represents the bare phonon Green's function.  It can be proved  that $\Sigma^{(1)}$ and $\Gamma^{(1)}$ respect the Ward identity for  the $\q=0$ mode, viz. 
 $\Gamma^{(1)}_{\k,\k}(E,E+\omega) = [\Sigma^{(1)}_\k (E) - \Sigma^{(1)}_\k(E+\omega)]/ \omega$ \cite{SP-PRB-2024}. 

Now,  as required  by the Ward identity, the identical vanishing of the phonon self-energy $\Pi_\q(\omega)$ for the $\q=0$ mode for an arbitrary  value of $g$  demands  that  each term in the expansion of  $\Pi_\q(\omega)$ in Eq. (\ref {Eq-Phonon-SE-expansion}) must vanish  order-by-order for the $\q=0$ mode. This is indeed the case, as demonstrated explicitly in Ref.\cite{SP-PRB-2024} for both  $\Pi^{(1)}$ and $\Pi^{(2)}$. While the leading-order phonon self-energy vanishes identically, the subleading-order phonon self-energy  vanishes 
because  the   contributions from the  vertex corrections cancel exactly those from the electron self-energy,
 i.e.,  $\Pi^{(2a)}_0(\omega) +\Pi^{(2b)}_0(\omega) =- \Pi^{(2c)}_0(\omega)$. Furthermore, the exact cancellation holds irrespective of the microscopic details of the system such as adiabaticity, dimensionality, etc. 

In the followings we extend our analysis for the specific case of the polaron in one-dimension in order to understand the momentum-dependent evolution of the impact of the many-body effects on the phonon spectral function of the polaron.

\subsection { Phonon self-energy in case of polaron }

\begin{figure}
\begin{center}
\hspace*{-8mm}
\resizebox{110mm}{115mm}{\psfig{figure=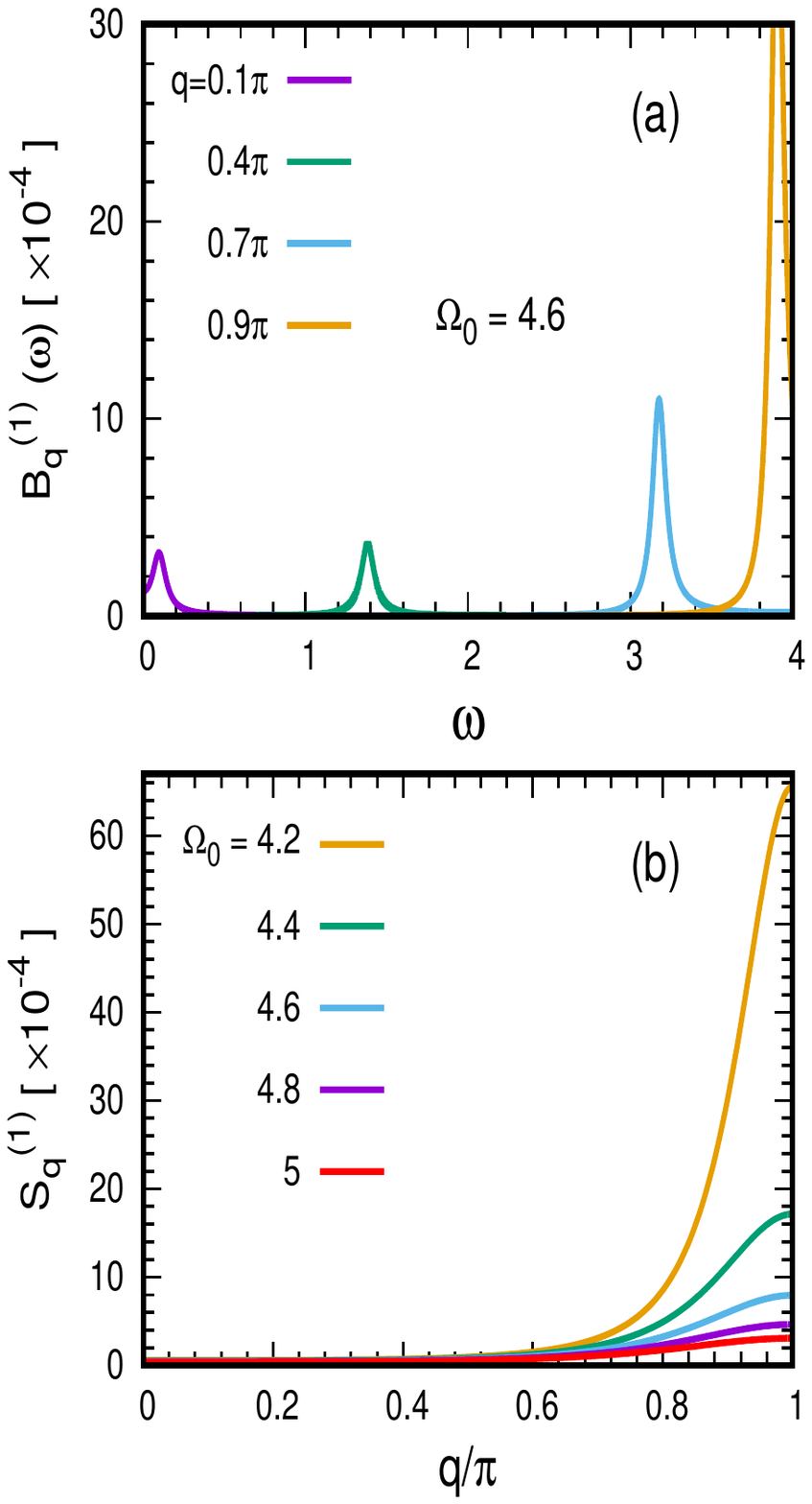}}
\end{center}
\vspace*{-5mm}
\caption{(a) Emergence of the polaronic features in the phonon spectral function due to interaction between the bare electron and  phonons, as demonstrated for $\Omega_0=4.6$ and four different values of $\q$. (b) The spectral weight [$S_\q^{(1)}$] associated with the polaronic features,  which are  peaked at the quasiparicle  energy $\omega=\epsilon_\q$, increases with $\q$ as $[\epsilon_\q^2-\Omega_0^2]^{-2}$ and therefore becomes more pronounced near the zone-boundary, as demonstrated for different values of $\Omega_0$. These results have been obtained with parameters $g=0.5,  N=1000$. }
\label{Fig-B1}
\end{figure}

In general, the leading- and subleading-order phonon self energies, as discussed above, can be evaluated for the case of an arbitrary band  dispersion ($\epsilon_\k$) and band filling, which determines the Fermi energy ($\epsilon_F$),  by substituting  $G^0_\k(E)=\frac{1-n_\k}{E-\epsilon_\k + i\eta} +  \frac{n_k}{E-\epsilon_\k - i\eta}$, where $n_\k=\theta( \epsilon_F -  \epsilon_\k )$, and  $D^{0}_\q (\omega)= \frac{1}{\omega-\Omega_0 + i\eta} - \frac{1}{\omega+\Omega_0 - i\eta}$. For the sake of quantitative investigations   we restrict our analysis to the case of a one-dimensional lattice by considering the nearest-neighbor tight-binding model for the band dispersion  $\epsilon_\k = 2t [1-\cos (\k a)]$,  where $t$ and $a$ represent hopping parameter and lattice constant, respectively. For the quantitative investigations that follow, all energies are measured in units of $ t $ while the wave-vectors are measured in units of the reciprocal lattice constant $(1/a)$. For all the quantitative results, as disussed in the followings,  we have used the scales where $t=1, a=1, \hbar =1$.  

The specific case of the single polaron can be analyzed by setting $n_\k=\delta_{\k,0}$  in $G^0_\k(E)$ corresponding to the lowest occupied Bloch state by the single electron  in the ground state. Now we evaluate the phonon self-energy in presence of the polaron as follows.  First we consider the leading-order phonon self-energy $\Pi_\q^{(1)}(\omega)$  in Eq. (\ref{Eq-Pi1}).  Substituting  the bare Green's function $G^0$ for the single electron and carrying out the integration over the energy variables analytically using the standard methods of contour integration,  the leading-order phonon self-energy can be expressed as follows.

\bea
\Pi_\q^{(1)}(\omega) &=& -\frac{g^2}{N}\sum_\k \left[    \frac{\delta_{\k+\q,0}(1-\delta_{\k,0})}{\omega+\epsilon_\k-\epsilon_{\k+\q}-i\eta} -   \frac{\delta_{\k,0}(1-\delta_{\k+\q,0})}{\omega+\epsilon_\k-\epsilon_{\k+\q}+i\eta}\right]\nonumber\\ \nonumber \\
&=& 
 -\frac{g^2}{N} \left \{ \frac{1}{\omega + \epsilon_\q-i\eta} - \frac{1}{\omega - \epsilon_\q +i\eta} \right \} (1-\delta_{\q,0}) \;.
\label{Eq-Pi1-p}
\eea
\noindent 
As expected,  we find that  $\Pi_\q^{(1)}(\omega)=0$ for the $\q=0$ mode. We note that while the leading-order phonon self-energy has  structure similar to that of  the bare phonon Green's function $D^0$, however, it has an explicit momentum dependence where the dispersionless  bare phonon frequency ($\Omega_0$) in $D^0$  is replaced by the quasiparticle dispersion ($\epsilon_\q$). This momentum dependence of $\Pi^{(1)}$ gives rise to appearance of  polaron-induced characteristic features  in phonon  spectral function, as discussed below in Sec. III. 

Similarly, we evaluate the different contributions of the subleading-order phonon self-energy $\Pi^{(2)}$ as follows.  We note that while the leading-order phonon self-energy  involves only $O(1/\mbox N)$  contribution, as can be seen from Eq. (\ref{Eq-Pi1-p}),  it turns out that the different components  of the subleading-order phonon self-energy, viz. $\Pi^{(2a)}, \Pi^{(2b)}, \Pi^{(2c)}$,  involve both $O(1/\mbox N)$ contributions, which include terms of type $\frac {1}{\mbox N^2}  \sum_{\k,\k'} (1-\delta_{\k,0}) \delta_{\k',0} \{   \}$,  as well as $O(1/\mbox N^2)$ contributions, which include terms of type  $\frac {1}{\mbox N^2} \sum_{\k,\k'} \delta_{\k,0} \delta_{\k',0} \{   \}$. Here we note that  $\sum_{\k} (1-\delta_{\k,0})  \{   \}  \sim O({\mbox N})$ and $\sum_{\k} \delta_{\k,0}  \{   \} \sim O(1)$. Retaining only the $O(1/{\mbox N})$ terms, which give the dominant contributions  in case of large $\mbox N$ in the thermodynamic limit, the different contributions of the subleading-order phonon self-energy can be expressed as follows. 

\begin{widetext}
\bea
\Pi^{(2a)}_\q(\omega)   
&=&-i \frac{g^2}{N} \sum_\k \int \frac{d E}{2\pi}  [G^0_\k(E)]^2 \Sigma^{(1)}_\k(E)G^0_{\k+\q}(E+\omega)
\nonumber \\ 
 &=&  \frac { g^4}{N^2} \sum_{\k} (1-\delta_{\k,0})  \left [  \left \{- \frac{(1 -\delta_{\k+\q,0})} { (\epsilon_{\k}+\Omega_0-  i\eta)^2 (\epsilon_{\k+\q}+\Omega_0- \omega - i\eta)}
  + \frac{1 }{ (\epsilon_\k+\Omega_0-  i\eta) (\epsilon_\q- \omega - i\eta)} \left (  \frac{1 }{ \epsilon_{\k}+\Omega_0-  i\eta}  + \frac{1} {\epsilon_\q- \omega - i\eta} \right  ) 
 \right. \right. \nonumber \\
&-& \left. \left.
 \frac{1 }{ (\epsilon_\q+\omega-  i\eta)^2 (\epsilon_\k +\Omega_0 + \omega - i\eta)}
 \right \} (1-\delta_{\q,0})
-\frac{2\delta_{\q,0}}{(\epsilon_\k +\Omega_0 -i\eta) \{ (\epsilon_\k +\Omega_0 -i\eta)^2 -\omega^2 \} }  
 \right ] 
 \label{Eq-pi2a-p}
\eea

\bea
 \Pi^{(2b)}_\q(\omega) &=& -i \frac{g^2}{N} \sum_\k \int \frac{d E}{2\pi}G^0_\k(E) [G^0_{\k+\q}(E+\omega)]^2 \Sigma^{(1)}_{\k+\q}(E+\omega) 
\nonumber \\
&=& \Pi^{(2a)}_{-\q}(-\omega)
\label{Eq-pi2b-p}
\eea

 \bea
  \Pi^{(2c)}_\q(\omega)  &=&-i \frac{g^2}{N} \sum_\k \int \frac{d E}{2\pi} 
G^0_\k(E)G^0_{\k+\q}(E+\omega)
\Gamma^{(1)}_{\k,\k+\q}(E,E+\omega)
\nonumber \\
&=& \frac { 2g^4}{N^2} \sum_{\k}  (1-\delta_{\k,0}) (1-\delta_{\k+\q,0})  \left [ - (1-\delta_{\q,0})  \left \{ \frac{1} 
{(\epsilon_\q- \omega - i\eta)(\epsilon_{\k+\q}+\Omega_0 -\omega -  i\eta) (\epsilon_\k+\Omega_0- i\eta)}
 \right. \right. \nonumber \\
&+& \left. \left.
\frac{1}{(\epsilon_\q+ \omega - i\eta)(\epsilon_\k+\Omega_0 +\omega -  i\eta) (\epsilon_{\k+\q}+\Omega_0- i\eta)} \right \}
+\frac{2\delta_{\q,0}} {  (\epsilon_\k+ \Omega_0 - i\eta) \{ (\epsilon_\k+\Omega_0  -  i\eta)^2 - \omega^2 \} }
  \right ]
 \label{Eq-pi2c-p}
 \eea
\end {widetext}

\noindent Here we have separated out the contributions for the $\q=0$ mode and the finte-$\q$ modes.  It must be emphasized that for the $\q=0$ mode although the many-body correction to the phonon self-energy involves  finite contributions from  both the electron-self energy,  as can be seen from Eqs. (\ref {Eq-pi2a-p}), (\ref {Eq-pi2b-p}),  as well as the vertex corrections, as can be seen from  Eq. (\ref {Eq-pi2c-p}), it turns out that the net contributions  from the electron self-energy is cancelled exactly by those of the vertex corrections,  i.e.,  $\Pi^{(2a)}_{\q=0}(\omega) + \Pi^{(2b)}_{\q=0}(\omega) =-\Pi^{(2c)}_{\q=0}(\omega)$, as required by the charge-conserving Ward identity.

\section{Polaron induced features in phonon spectral function}
We now  exploit the results for the phonon self-energy, as discussed above, for analyzing the phonon spectral function $B_\q(\omega)=- \frac{1}{\pi} \mbox{Im} D_\q (\omega)$ of the polaron, where  $D$ represents the renormalized phonon Green's function. 
Using the  Dyson equation, $D_\q(\omega) = D_\q^0 (\omega) + D_\q^0 (\omega)\Pi_\q(\omega)D_\q(\omega)$, the renormalized phonon Green's function in the weak-coupling limit  can be approximated  by retaining  only the term representing the leading many-body correction, i.e.,  $D_\q(\omega) \approx D_\q^0 (\omega) + D_\q^0 (\omega)\Pi_\q(\omega)D_\q^0(\omega)$.  Now, considering the phonon self-energy by taking into account the leading- and subleading-order contributions of 
 $O(1/\mbox N)$, i.e., $\Pi \approx \Pi^{(1)} + \Pi^{(2)}$, as discussed above, the phonon Green's function of the interacting system can  be expanded as 
 $D=D^0 + D^{(1)} + D^{(2)}$. Here $D^{(1)}=[ D^0]^2 \Pi^{(1)}$ and $D^{(2)}=[ D^0]^2 \Pi^{(2)}$ represent the leading- and subleading-order corrections of  $O(1/\mbox N)$ to the phonon Green's function, respectively. Therefore,  the phonon spectral function  of the interacting system can be written as $B_\q (\omega)  = B^0_\q(\omega) + B^{(1)}_\q(\omega)+ B^{(2)}_\q(\omega)$. Here $B^0_\q(\omega) = -\frac{1}{\pi}\Im D^0_\q(\omega)= [\delta(\omega-\Omega_0) +  \delta(\omega+\Omega_0)]$ represents the spectral function in case of the non-interacting $(g=0)$ system which is characterized by a single peak at $|\omega|=\Omega_0$ representing the dispersionless optical phonons of frequency $\Omega_0$.  On the other hand, $B^{(1)}=-\frac{1}{\pi}\Im D^{(1)}_\q(\omega) $ and $B^{(2)}=-\frac{1}{\pi} \Im D^{(2)}_\q(\omega)$ provide the fingerprints of the additional features which appear in the phonon spectral function owing to the electron-phonon coupling induced polaron formation.

In order to simplify our analysis of the polaronic  features in the  phonon spectral function we focus on  the low-energy regimes $|\omega| <  \Omega_0$ 
where $ \Im D^0_\q(\omega)=0$. Therefore, the leading- and subleading-order corrections to the phonon spectral function can be written as $B^{(n)} = -\frac{1}{\pi} \Im [\Pi^{(n)}_\q(\omega ) \{D_\q^0(\omega) \}^2]=  -\frac{1}{\pi} [\Im \Pi^{(n)}_\q(\omega )] \{ \Re D_\q^0(\omega) \}^2= -\frac{1}{\pi} [\Im \Pi^{(n)}_\q(\omega )]  \frac{4\Omega_0^2}{(\omega^2-\Omega_0^2)^2} $, where $n=1,2$. Now we analyze the imaginary part of the leading- and subleading-order phonon self-energy for understanding the emergence of  the low-energy polaronic features in the phonon spectral function, as follows.

First we consider the leading-order phonon self-energy $\Pi^{(1)}_\q(\omega)$ as given in  Eq. (\ref {Eq-Pi1-p}).  Extracting out its imaginary part, we get 

\bea
  \Im \Pi^{(1)}_\q (\omega) &=& -\frac { g^2}{N} \pi \left \{ \delta (\omega -\epsilon_\q)  + \delta (\omega+\epsilon_\q) \right \}  \; .
\label{Eq-impi1}
\eea 

\noindent Therefore,  the leading-order correction to the phonon spectral function owing to coupling between the bare electron and phonons, as represented by $B_\q^{(1)}(\omega) =  -\frac{1}{\pi} [\Im \Pi^{(1)}_\q(\omega )]  \frac{4\Omega_0^2}{(\omega^2-\Omega_0^2)^2}$, can be written as,

\bea
B_\q^{(1)}(\omega) =S^{(1)} (\omega) \left \{ \delta (\omega - \epsilon_\q) +   \delta (\omega + \epsilon_\q)  \right \}  \:,
\label {Eq-B1}
\eea

\noindent where $S^{(1)}(\omega)= \frac { g^2}{N} \frac{4\Omega_0^2}{(\omega^2-\Omega_0^2)^2}$. As obvious from the structure of $B^{(1)}$, the coupling between bare electron and phonons results in emergence of  the polaron-induced local peaks in the phonon spectral function at the quasiparticle energy $|\omega| = \epsilon_\q$, as demonstrated in Fig. (\ref {Fig-B1}) for different  values of $\q$.  Here we have shown the results only  for $\omega > 0$ because $B_\q^{(1)}(\omega)$ is symmetric with respect to $\omega$.  With increasing $\q$ the strength  of the spectral weight $S^{(1)}_\q = S^{(1)}(\omega=\epsilon_\q)$ associated with the polaronic peaks  increases 
as  $[\epsilon_\q^2-\Omega_0^2]^{-2}$.

Next, we analyze the subleading-order phonon self-energy $\Pi^{(2)}_\q(\omega)$ for understanding the impact of  the many-body effects on the low energy polaronic features, as described  by $B_\q^{(2)}(\omega) =  -\frac{1}{\pi} [\Im \Pi^{(2)}_\q(\omega )]  \frac{4\Omega_0^2}{(\omega^2-\Omega_0^2)^2}$.  We evaluate the different components of $\Im \Pi^{(2)}$ using  Eqs. (\ref{Eq-pi2a-p}),  (\ref{Eq-pi2b-p}), and (\ref{Eq-pi2c-p}), as follows.  Firstly, we note that  there is no contribution from the imaginary  part of the $\omega-$independent term $1/(\epsilon_\k+\Omega_0-  i\eta)$ so we set $1/(\epsilon_\k+\Omega_0-  i\eta)=1/(\epsilon_\k+\Omega_0)$. Secondly, since $\epsilon_\k  [ = 2(1-\cos\k)] \ge 0$, therefore in the low energy regime [$ |\omega| < \Omega_0$]  there is no contribution to the imaginary parts of $\Pi^{(2a)}_\q(\omega ), \Pi^{(2b)}_\q(\omega )$ and $\Pi^{(2c)}_\q(\omega )$  from  the following $\omega-$dependent terms: (i) $1/(\omega-\epsilon_{\k+\q}-\Omega_0+ i\eta)$, and (ii) $1/(\omega+\epsilon_\k +\Omega_0  - i\eta)$. Therefore, we set $ 1/(\omega-\epsilon_{\k+\q}-\Omega_0+ i\eta)=1/(\omega-\epsilon_{\k+\q}-\Omega_0)$, and  $1/(\omega+\epsilon_\k +\Omega_0  - i\eta) = 1/(\omega+\epsilon_\k +\Omega_0)$. For separating out the real and imaginary components from the remaining terms  
we use the  relations $\frac{1}{\omega - \xi \pm i\eta} = P \frac{1}{(\omega-\xi)} \mp i\pi \delta(\omega-\xi)$ and $\frac{1}{(\omega - \xi \pm i\eta)^2} = P \frac{1}{(\omega-\xi)^2} \mp i\pi \delta(\omega-\xi) \frac {d}{d\omega}$ \cite{Levine-JRNIST-2008}. Following these steps  
we obtain the imaginary part of  the different components of  $\Pi^{(2)}$ as follows.  Representing the net contributions from the electron self-energy as $\Pi^{(2ab)}$, i.e., $\Pi^{(2ab)} = \Pi^{(2a)} + \Pi^{(2b)}$,  we get
\bea
\Im \Pi^{(2ab)}_\q(\omega) & = & \frac{g^2}{N} \pi \left \{  f_\q^{(2ab)}(\omega)   \delta (\omega - \epsilon_\q)
\right. 
\nonumber \\
& + & \left.  f_{-\q}^{(2ab)}(-\omega)   \delta (\omega + \epsilon_\q)  \right \} \;,
\eea
\noindent where,

\bea
 f_\q^{(2ab)}(\omega)  
=\frac { g^2}{N}
\sum_{\k}    \left \{
 \frac{1-\delta_{\k,0}}{(\epsilon_\k+\Omega_0)^2}
+\frac{1-\delta_{\k+\q,0}} {(\epsilon_{\k+\q}+\Omega_0-\omega)^2}  \right   \}   \;.
 \label{Eq-fq2ab}
\eea

Similarly, the contributions from the vertex corrections can be expressed as,
\bea
\Im \Pi^{(2c)}_\q(\omega) & = & \frac{g^2}{N} \pi \left \{  f_\q^{(2c)}(\omega)   \delta (\omega - \epsilon_\q) 
\right . \nonumber \\
&+ & \left.   f_{-\q}^{(2c)}(-\omega)   \delta (\omega + \epsilon_\q)  \right \} \;,
\eea
\noindent where,
\bea
 f^{(2c)}_\q(\omega)  
=-\frac {2g^2}{N}
\sum_\k  \frac {(1-\delta_{\k,0})  ( 1-\delta_{\k+\q,0 })} {(\epsilon_{\k+\q} +\Omega_0-\omega)(\epsilon_\k+\Omega_0)}  \;.
 \label{Eq-fq2c}
\eea

Therefore, the net many-body correction  to the phonon self-energy $\Im \Pi^{(2)}_\q(\omega)=\Im \Pi^{(2ab)}_\q(\omega) +\Im \Pi^{(2c)}_\q(\omega)$
can be written as, 
\bea
\Im \Pi^{(2)}_\q(\omega) & = & \frac{g^2}{N} \pi \left \{  f_\q^{(2)}(\omega)   \delta (\omega - \epsilon_\q) +  f_{-\q}^{(2)}(-\omega)   \delta (\omega + \epsilon_\q)  \right \} \;,
\nonumber \\
\label{Eq-impi2net}
\eea
\noindent where,

\bea
 f^{(2)}_\q(\omega)
&=&  f_\q^{(2ab)}(\omega) + f_\q^{(2c)}(\omega)
\nonumber\\
&=& 
\frac { g^2}{N}
\sum_{\k}   \left \{  \frac{1-\delta_{\k,0}}{\epsilon_{\k}+\Omega_0 }-  \frac{ 1-\delta_{\k+\q,0} }{\epsilon_{k+q}+\Omega_0 -\omega } \right \}^2 
 \label{Eq-fq2}
\eea

\noindent  Now, the net many-body correction to the phonon spectral function represented by  $B_\q^{(2)}(\omega) =  -\frac{1}{\pi} [\Im \Pi^{(2)}_\q(\omega )]  \frac{4\Omega_0^2}{(\omega^2-\Omega_0^2)^2}$,  as discussed above, can be expressed as  
\bea
B_\q^{(2)}(\omega)  &=&-S^{(1)} (\omega)  \left \{ f_\q^{(2)} (\omega)  \delta (\omega - \epsilon_\q) +  f_{-\q}^{(2)} (-\omega)  \delta (\omega + \epsilon_\q)  \right \} 
\nonumber \\
\; .
\label {Eq-B2}
\eea

\begin{figure}
\vspace*{0mm}
\begin{center}
\hspace*{0mm}
\resizebox{85mm}{120mm}{\psfig{figure=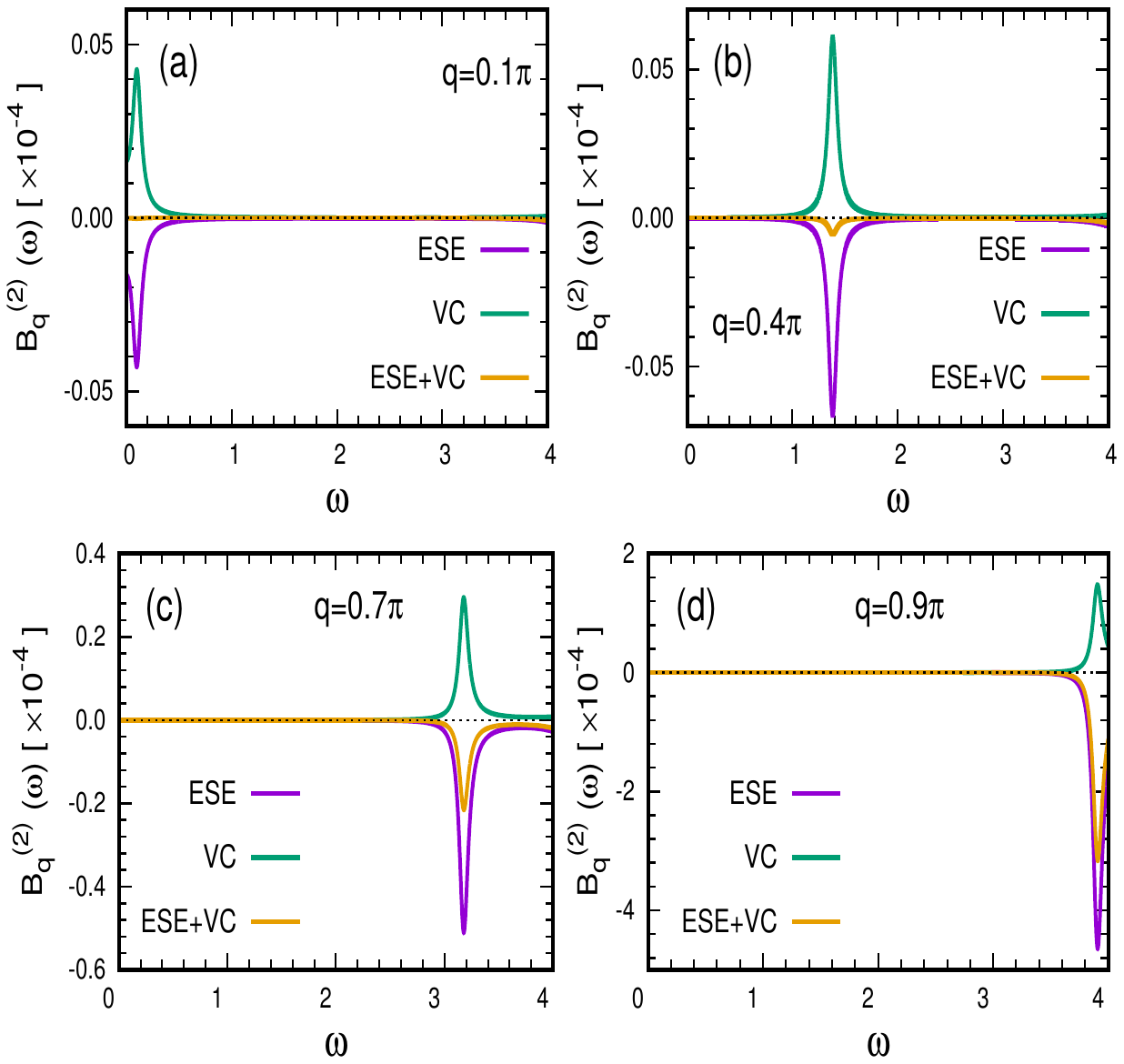}} 
\end{center}
\vspace*{-40mm}
\caption{Contributions from electron self-energy (ESE) and vertex corrections (VC)  result in suppression and enhancement in the spectral weight of the polaronic features in the phonon spectral function, respectively,   as demonstrated for four different values of $\q$.  The two opposite contributions nearly cancel each other for the small $\q$ modes, as can be seen for $\q=0.1\pi$  (a) and  $0.4\pi$ (b), thereby leaving the polaronic spectral weight nearly unaffected by the many-body effects. With increasing
 $\q$, while strength of both the contributions increases,  the net many-body correction is dominated by the rapid increase in the contribution from electron self-energy thereby resulting in a finite suppression of the spectral weight which becomes more pronounced near the zone-boundary (c,d). These results have been obtained with parameters $g=0.5, \Omega_0=4.6, N=1000$.}
\label{Fig-B2}
\end{figure}

\noindent In the followings we focus our analysis  only in the region $\omega >0$ as $B^{(2)}_\q(\omega)$ is symmetric in $\omega$.  As evident from Eqs. (\ref{Eq-B1}) and (\ref{Eq-B2}), $B^{(2)}$ can also be written as $B^{(2)}_\q(\omega) = -f_\q^{(2)} (\omega) B^{(1)}_\q(\omega) $, where $B^{(1)}$ represents the polaronic features in phonon spectral function arising due to interaction between the bare electron and phonons.  Similarly,  the many-body corrections to the phonon spectral function arising due to contributions from the electron self-energy, as denoted by 
 $B^{(2ab)}_\q(\omega) =   -\frac{1}{\pi} [\Im \Pi^{(2ab)}_\q(\omega )]  \frac{4\Omega_0^2}{(\omega^2-\Omega_0^2)^2}$,  and the vertex corrections, as denoted by   $B^{(2c)}_\q(\omega) =  -\frac{1}{\pi} [\Im \Pi^{(2c)}_\q(\omega )]  \frac{4\Omega_0^2}{(\omega^2-\Omega_0^2)^2}$, can also be related to $B^{(1)}$ as $B^{(2ab)}_\q(\omega) = -f_\q^{(2ab)} (\omega) B^{(1)}_\q(\omega) $ and $B^{(2c)}_\q(\omega) = -f_\q^{(2c)} (\omega) B^{(1)}_\q(\omega) $, respectively,  such that the net many-body correction becomes $B^{(2)}_\q(\omega) = B^{(2ab)}_\q(\omega) +B^{(2c)}_\q(\omega)$.

Physically,  the parameters $f_\q^{(2ab)}(\omega)$ and $f_\q^{(2c)}(\omega)$ provide a measure of  strength of the many-body correction  arising due to  contributions from the electron self-energy and vertex corrections,  respectively, while $f_\q^{(2)}(\omega) [= f_\q^{(2ab)}(\omega) + f_\q^{(2c)}(\omega)] $ represents their net impact. We find that  $f_\q^{(2ab)}(\omega) >0$, as evident from Eq. (\ref{Eq-fq2ab}),  so that   $B^{(2ab)}_\q(\omega) = -f_\q^{(2ab)} (\omega) B^{(1)}_\q(\omega)  < 0$, which represents suppression in spectral weight of the polaronic features. On the other hand,  as evident from Eq.  (\ref{Eq-fq2c}),  we find that $f_\q^{(2c)}(\omega) < 0$, so that $B^{(2c)}_\q(\omega) = -f_\q^{(2c)} (\omega) B^{(1)}_\q(\omega)  >0$, which represents enhancement  in the spectral weight of the polaronic features.  In other words, the vertex corrections always tend to compensate the impact of the electron self-energy. Furthermore, as demonstrated in Fig. (\ref{Fig-B2}) for different $\q$ values,  the impact of both the  electron self-energy as well as the  vertex corrections becomes more pronounced with increasing $\q$.   For the net many-body correction, we find that $f^{(2)}_\q(\omega) \ge 0$, as evident from Eq. (\ref{Eq-fq2}),   so that  $B^{(2)}_\q(\omega) = -f_\q^{(2)} (\omega) B^{(1)}_\q(\omega)   \le 0$  where the equality sign holds strictly for the $\q=0$ mode. In other words, while the  contributions from the vertex corrections and  electron self-energy  are always of opposite nature, they cancel each other exactly only for the $\q=0$ mode thereby resulting in the identical vanishing of the net many-body correction as a consequence of the charge-conserving Ward identity. For $\q > 0$ modes, the net many-body correction  becomes finite, viz.  $f^{(2)}_\q(\omega) > 0$, which is of the same sign as that of the contribution from the electron self-energy. Therefore, the net many-body correction  for $\q>0$ modes is dominated  by the contribution from the electron self-energy which results in a net suppression of the polaronic spectral weight,  as shown in Fig. (\ref{Fig-B2}). With increasing $\q$ the impact of many-body effects becomes more pronounced, as apparent from the strong suppression in the polaronic spectral weight for the  $\q$ modes near the zone boundary in  Fig. (\ref{Fig-B2}).

\begin{figure}
\vspace*{0mm}
\begin{center}
\hspace*{-15mm}
\resizebox{110mm}{120mm}{\psfig{figure=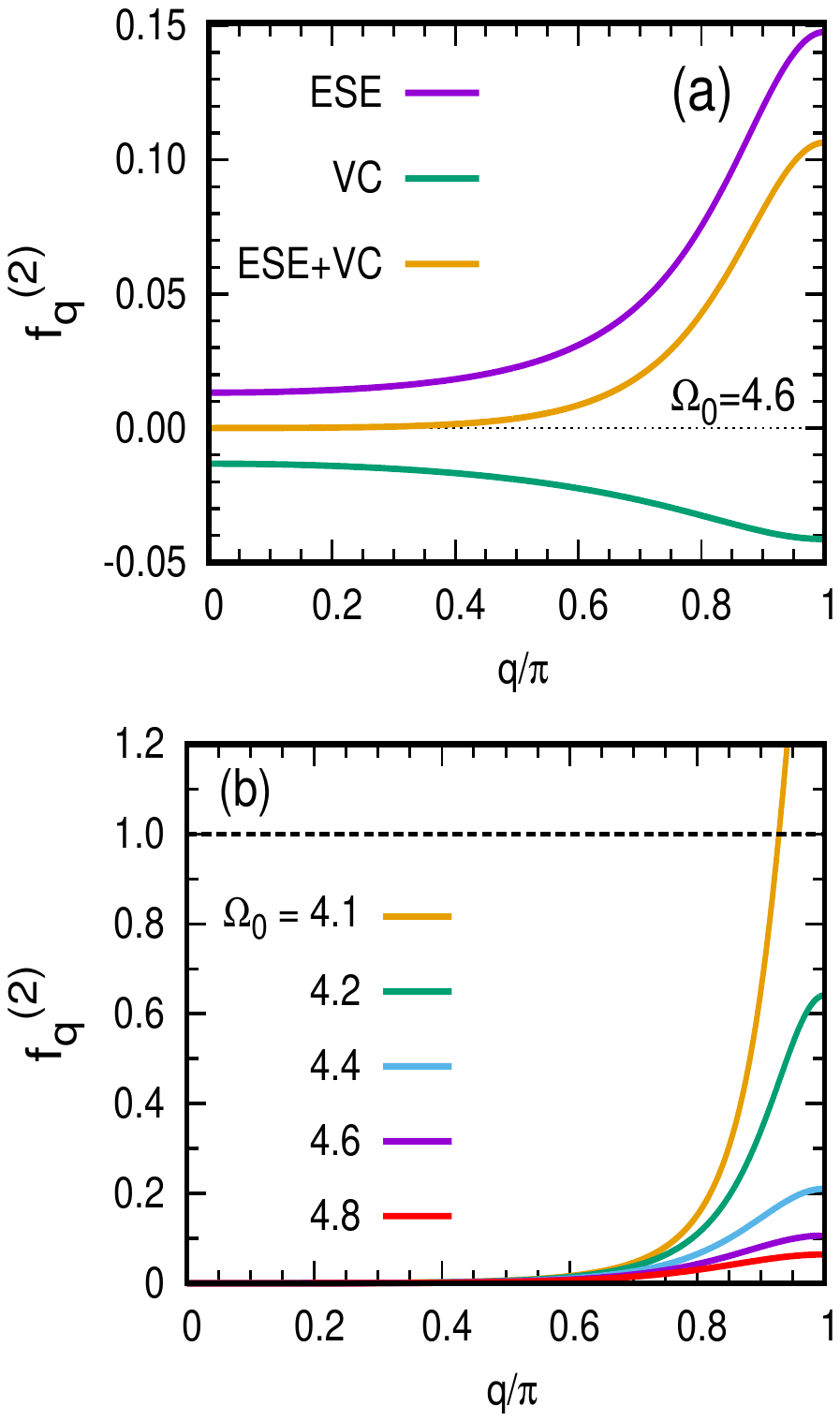}}
\end{center}
\vspace*{-20mm}
\caption{(a) Momentum dependent evolution of the strength of the many-body correction to the polaronic spectral weight arising due to the contributions  from electron self-energy (ESE) and vertex corrections (VC), as shown for $\Omega_0=4.6$. Being of the opposite signs the two contributions nearly cancel each other for the small-$\q$ mode which  results in a negligible impact of the many-body effects.  On the other hand, for large-$\q$ modes, the many-body correction is dominated by the contribution of electron self-energy due to its rapid increase in comparison to that of the vertex correction, thereby  resulting in a finite suppression of the spectral weight which becomes more pronounced near the zone-boundary. (b) The net many-body correction increases with  decreasing $\Omega_0$  because the rapid increase in the contribution of electron self-energy for the large $\q$ modes, as shown above in (a),  becomes more pronounced  with  decreasing $\Omega_0$. The crossover $f_\q^{(2)} > 1$ in the limiting case $\Omega_0 \rightarrow W$, as demonstrated  for $\Omega_0=4.1$, represents the breakdown of our perturbative approach, as discussed in the text.  These results have been obtained with parameters $g=0.5, N=1000$.}
\label{Fig-fq2}
\end{figure}

In order to further elaborate the momentum-dependent evolution of  the impact of the many-body effects we introduce the parameters $f_\q^{(2ab)}= f_\q^{(2ab)} (\omega=\epsilon_\q)$ and $f_\q^{(2c)}= f_\q^{(2c)} (\omega=\epsilon_\q)$ which  represent the renormalization factors by which the polaronic spectral weight   for a particular $\q$ mode is suppressed and enhanced due to the contributions from the electron self-energy and vertex corrections, respectively. As demonstrated in Fig. (\ref{Fig-fq2}), with increasing $\q$,  while  strength of both the contributions  increases,  the contribution from electron self-energy increases faster than that of the vertex corrections which becomes more pronounced near the zone-boundary.  In other words, despite the two opposite contributions, a finite net many-body correction 
as represented by the parameter $f_\q^{(2)}=f_\q^{(2ab)} +f_\q^{(2c)}$,  arises  due to dominance of the contributions from the electron self-energy, specifically near the zone-boundary. Furthermore, as demonstrated in Fig, (\ref{Fig-fq2}), with decreasing $\Omega_0$ the net many body correction for the large-$\q$ modes shows substantial enhancement because the rapid increase in the  contribution from the electron self-energy becomes more pronounced.  However, in the regime of long-wavelength excitations the  net many-body correction continues to  remain small due to nearly equal strength of the  two opposite contributions, which can also be naively expected from their continuous evolution from the $\q=0$ mode where they cancel each other exactly. Our results for the enhancement in the impact of the many-body effects with increasing momentum are also relevant for a qualitative understanding of the strong renormalization of the zone-edge modes in comparison to that of the zone-center modes, as observed recently in case of the layered compound WS$_2$ using the Raman spectroscopy\cite{Wang-PRB-2024}, in terms of the momentum dependent evolution of competing contributions from the electron self-energy and vertex corrections, as discussed above. Similar momentum-dependent renormalization of the phonon spectrum has also been reported earlier in case of the monolayer graphene\cite{Dresselhaus-PRL-2012}.

The quantative results demonstrated in Figs. (\ref{Fig-B2}) and (\ref{Fig-fq2}) highlight  the importance of the vertex corrections for a proper estimation of the impact of the many-body effects on the phonon spectral function associated with the polaronic features.  
While spectral weight of the  polaronic features is always suppressed due to the contribution from the electron self-energy, the compensating contribution from the vertex corrections tends  to restore the spectral weight thereby weakening the  impact of the many-body effects.  Indeed, due to a near cancellation between the two contributions the polaronic spectral weight remains almost unaltered from the impact of the many-body effects  for the small-$\q$ modes. Although the contribution of electron self-energy becomes dominant for the large-$\q$ modes near the zone-boundary, the compensating contribution from the vertex corrections still remains significant for an accurate estimation of the impact of many-body effects.

It is worth mentioning the limitation of our weak-coupling analysis. As $B_\q^{(0)}=0$ for $\omega < \Omega_0$, the phonon spectral function describing the low-energy  polaronic features can be written as $B_\q(\omega)= B_\q^{(1)}(\omega) +  B_\q^{(2)}(\omega)= B_\q^{(1)}(\omega) [1 -  f_\q^{(2)}(\omega)]=S_\q(\omega)\delta(\omega-\epsilon_\q)$, where $S_\q(\omega)=  S^{(1)}(\omega) [1 -  f_\q^{(2)}(\omega)]$, so that $S_\q=S_\q(\omega=\epsilon_\q)$ provides a measure of the net polaronic spectral weight in the phonon spectral function. Here $f_\q^{(2)} (\ge 0)$ represents the renormalization factor by which the spectral weight associated with the polaronic features is suppressed due to the  many-body effects, as discussed above. Therefore our analysis is valid only in the parameter space where $f_\q^{(2)} < 1$.   While this condition is satisfied in the antiadiabatic regime for the small $\q$ modes, however, for the zone-boundary modes $f_\q^{(2)}$ increases rapidly with decreasing $\Omega_0$ and even exceeds '1' when $\Omega_0 \rightarrow W$, as also demonstrated  in Fig. (\ref{Fig-fq2}),  thereby invalidating our weak-coupling perturbative analysis of the many-body effects.   In other words, our estimation of the impact of the many-body effects for the zone-boundary modes becomes less accurate  for $\Omega_0 \rightarrow W$ when $f_\q^{(2)} \rightarrow 1$.  As discussed in the followings, the renormalization of the quasiparticle spectrum must be taken into account for an accurate estimation of the impact of the many-body effect for the large$-\q$ modes when $\Omega_0 \rightarrow W$ .

\section{Effect of renormalization of quasiparticle spectrum} 

Now we extend our analysis 
of the polaronic features in the phonon spectral function
by taking into account the renormalization of the  quasiparticle spectrum due to electron-phonon coupling. 
Upon carrying out the $E-$integration in Eq.( \ref{Eq-electron-SE})  for the specific case of the single polaron, it turns out that  the leading-order electron self-energy $\Sigma^{(1)}_\k(E)$ involves both $O(1)$ term, which corresponds to the contribution from the particle states, as well as  $O(1/\mbox N)$ term, which corresponds to the contribution from the single hole state. Ignoring the negligible   $O(1/\mbox N)$  
contribution,  the electron self-energy can be written as  
\bea
\Sigma_\k^{(1)} (E)
&=&  \frac {g^2}{N} \sum_\Q   \frac{1-\delta_{\k+\Q,0} }  {E-\Omega_0-\epsilon_{\k+\Q}+i\eta} \; .
\eea

\begin{figure}
\vspace*{-12mm}
\begin{center}
\hspace*{-5mm}
\resizebox{80mm}{100mm}{\psfig{figure=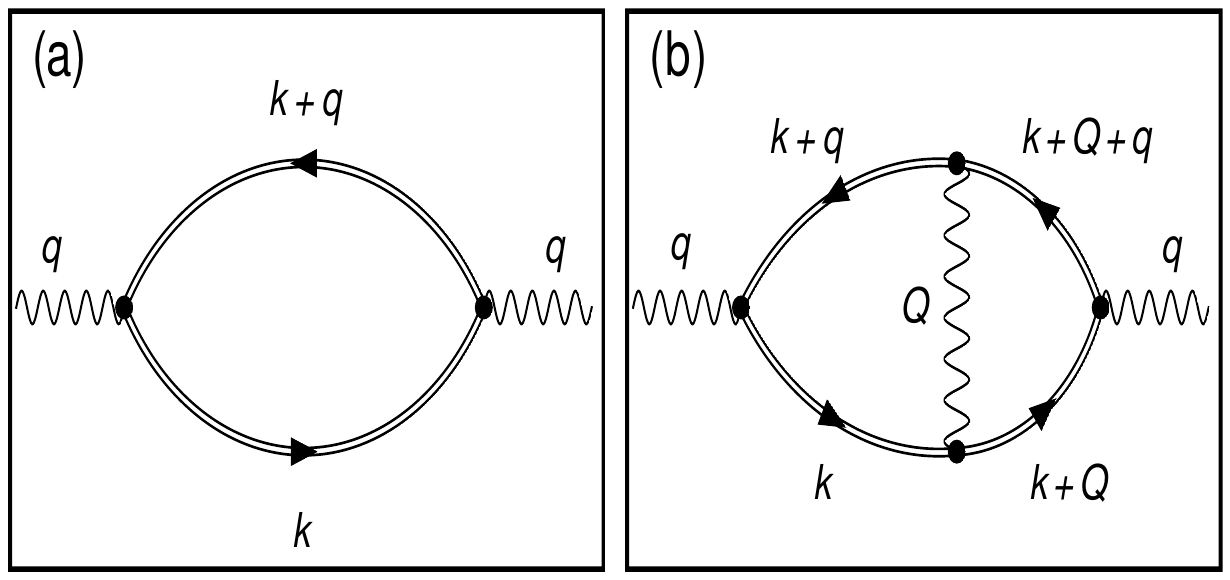}}
\end{center}
\vspace*{-50mm}
\caption{Diagrammatic representations of the phonon self-energies $\Pi^{(a)}$ (a) and $\Pi^{(b)}$ (b) where the double lines with arrow represent the dressed fermion propagator and the wavy lines represent the  bare phonon propagator.  While $\Pi^{(a)}$  represents only the impact of electron self-energy, on the other hand,  $\Pi^{(b)}$ reprsents the combined impact of both the electron self-energy and vertex corrections.  Here we follow the implicit notations $\q=(\q,\omega), \k=(\k,E), \Q=(\Q,\Omega)$.}
\label{Fig-Feynman-ren}
\end{figure}

\begin{figure}
\begin{center}
\hspace* {-20mm}
\resizebox{130mm}{130mm} {\psfig{figure=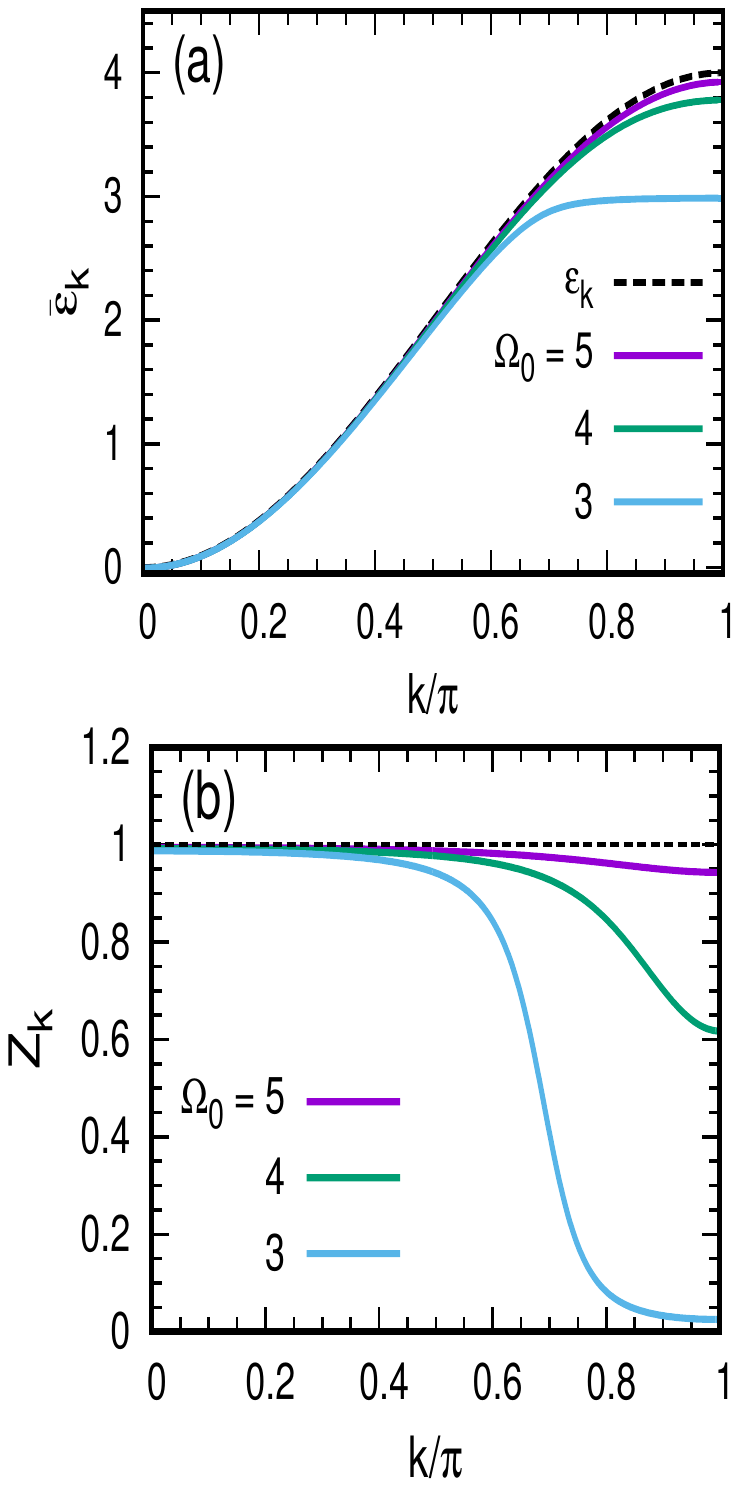}} 
\end{center}
\vspace*{-20mm}  
\caption{ Comparison of bare $(\epsilon_\k$) and renormalized $({\bar\epsilon_\k})$ quasiparticle dispersions (a) and momentum-dependent evolution of the renormalization factor $(Z_\k)$ (b), as demonstrated for different values of $\Omega_0$.  The deviation of $Z$ from its bare value '1' , as shown by the dotted line, provides a measure of the impact of the renormalization which weakens with increasing $\Omega_0$. In the antiadiabatic regime ($\Omega_0 > W$) the quasiparticle spectrum is weakly affected by the many-body effects as can be noticed from  $Z_k \approx 1$ and  ${\bar\epsilon_\k} \approx \epsilon_\k$ for $\Omega_0=5 $. These results have been obtained with parameters $g=0.5, N=1000$.}
\label{Fig-rdisp}
\end{figure}

\noindent 
We note that the electron self-energy is local, i.e., $\Sigma_\k^{(1)} (E)= \Sigma^{(1)} (E) $. 
For $E < \Omega_0$ we have  $\Im  \Sigma^{(1)} (E) =0$ so that  $\Sigma^{(1)} (E) =  \Re  \Sigma^{(1)} (E)$ which can be evaluated by carrying out the momentum summation analytically as 

 \begin{eqnarray}
\Re  \Sigma^{(1)} (E)
 &=& -\frac{g^2}{2t} \frac{1}{\sqrt{\alpha^2-1}} \; ,
 \end{eqnarray}

\noindent where, $\alpha=1+\frac{\Omega_0-E}{2t}$. Since  $\Re \Sigma^{(1)} (E) \rightarrow -\infty$ as $E\rightarrow \Omega_0$,  therefore, the renormalized quasiparticle dispersion (${\bar \epsilon_\k}$), which is obtained  self-consistently as 
 ${\bar \epsilon_\k}=\epsilon_\k + \Re \Sigma^{(1)} ({\bar \epsilon_\k})$, always remains smaller than $\Omega_0$, as demonstrated in Fig. (\ref{Fig-rdisp}) for different values of $\Omega_0$. Here we measure the electron self-energy relative to its value at $E=0$ so that $ {\bar \epsilon_\k}=0$ for $\k=0$. Therefore, the renormalization of the quasiparticle dispersion cures the issue of the unphysical divergence in the phonon spectral weight [ $S^{(1)}_\q$ ] of the polaron arising due to interaction between the bare electron and phonons for the specific $\q$ values  when the bare dispersion  $\epsilon_\q$ becomes equal to  $\Omega_0$, as discussed  above in Sec. III. 

Now, by taking into account the renormalization of quasiparticle spectrum and  approximating the vertex corrections by their leading-order contributions, i.e., 
$\Gamma \sim  1 + {\bar \Gamma}^{(1)}$,  the phonon self-energy, as defined in Eq.(\ref{Eq-phonon-SE}), can be written as $\Pi_\q(\omega)= \Pi^{(a)}_\q(\omega) +  \Pi^{(b)}_\q(\omega)$, where

\bea
\Pi^{(a)}_\q(\omega) &=&-i \frac{g^2}{N} \sum_\k \int \frac{d E}{2\pi} G_\k(E) G_{\k+\q}(E+\omega)
\label{Eq-Pia}
\eea
represents the impact of only the electron self-energy correction, and
\bea
\Pi^{(b)}_\q(\omega) &=&-i \frac{g^2}{N} \sum_\k \int \frac{d E}{2\pi} G_\k(E) G_{\k+\q}(E+\omega)  {\bar \Gamma}^{(1)}_{\k,\k+\q}(E,E+\omega)
\nonumber\\
\label{Eq-Pib}
\eea
represents the combined impact of both the electron self-energy and the vertex corrections. The diagrammatic representations for $\Pi^{(a)}$ and $\Pi^{(b)}$ are shown in Fig. (\ref{Fig-Feynman-ren}). By taking into account the renormalization of the quasiparticle dispersion, the leading-order vertex corrections can be expressed as

\bea
{\bar\Gamma}^{(1)}_{\k,\k+\q}(E,E+\omega) 
&=& i \frac{g^2}{N}
\sum_\Q \int \frac{d \Omega}{2\pi} G_{\k+\Q}(E+\Omega) 
\nonumber \\
& \times & 
G_{\k+\Q+\q}(E+\Omega+\omega) 
D^0_\Q(\Omega)
\;  , 
\eea

\noindent  Now, in order to simplify  our analysis we ignore the contribution of the high energy continuum so that the renormalized electron Green's function can be approximated in terms of its coherent part as 
\bea
G_\k(E)  & \approx & Z_\k  \left \{ \frac{1-\delta_{\k,0}} {E- {\bar \epsilon_\k}  +  i \eta } +  \frac{\delta_{\k,0} }{E-{\bar \epsilon_\k}- i \eta} \right \}.
\label {Eq-reGF}
\eea

 Here ${\bar \epsilon_\k}$ 
 represents the renormalized quasiparticle energy, as discussed above,  and  $Z_\k = \left \{ 1 - \frac{d}{dE} \Re \Sigma^{(1)} (E) \right \}^{-1}_{E ={\bar \epsilon_\k}}$ represents the renormalization factor.   The momentum-dependent variation of the renormalization factor for different values of $\Omega_0$ is shown in  Fig.  (\ref{Fig-rdisp}). The deviation of the renormalization factor from its bare value '1'  indicates the increasing impact of the many-body effects which results in increase in the  transfer of the electronic spectral weight into the high energy continuum.  Approximating  $G$  in terms of its coherent part becomes more accurate in cases when $Z_\k \approx 1$. This is indeed the case when the phonon energy $(\Omega_0)$ exceeds significantly the bare electronic bandwidth ($W$) in the antiadiabatic regime, as demonstrated for $\Omega_0=5$  in Fig. (\ref{Fig-rdisp}).  
 
Using the coherent part of the electron Green's function in Eq. (\ref {Eq-reGF})  the phonon self-energies   $\Pi^{(a)}$  and $\Pi^{(b)}$  can be  evaluated  by carrying out the integration over the energy variables analytically, as follows. First we consider the impact of electron self-energy, as described by $\Pi^{(a)}$, which can be expressed as 
\bea
\Pi^{(a)}_\q(\omega)&=& -\frac{g^2}{N}  Z_0 Z_\q  \left \{  \frac{1}{\omega + {\bar \epsilon_\q} -i\eta} - \frac{1}{\omega - {\bar \epsilon_\q} +i\eta} \right \}   (1-\delta_{\q,0}) \: .
\nonumber \\
\label{Eq-pia}
 \eea
As expected, $\Pi^{(a)}$ reduces to $\Pi^{(1)}$ in Eq. (\ref{Eq-Pi1-p}) in limits $Z_\q \rightarrow 1$ and ${\bar \epsilon_\q} \rightarrow \epsilon_\q$ which characterises the polaronic features arising due to interaction between the bare electron with the surrounding phonons.  

Next, we consider the combined impact of the electron self-energy and vertex corrections, as described by $\Pi^{(b)}$ which can be expressed as,

\bea
\Pi^{(b)}_\q(\omega)&=& \frac{g^2}{N}  Z_0 Z_\q  \left \{  \frac{ f_{-\q}(-\omega) } {\omega + {\bar \epsilon_\q} -i\eta} 
-  \frac{f_{\q}(\omega) } {\omega - {\bar \epsilon_\q} +i\eta}  \right \}   (1-\delta_{\q,0}),
\nonumber \\
\label{Eq-pib}
 \eea\

where, 

\bea 
 {f_{\q}(\omega)}
&=&
  -\frac{2g^2}{N} \sum_\k \frac{  (1-\delta_{\k,0})Z_{\k} (1-\delta_{\k+\q,0})Z_{\k+\q}} {({\bar \epsilon_{\k+\q}} + \Omega_0-\omega )
( {\bar \epsilon_{\k} + \Omega_0)}}
\label{Eq-Fq-renvertex}
\eea
We note that, unlike $f_\q^{(2c)}(\omega)$ in Eq. (\ref {Eq-fq2c}), which provides  a meaure of the impact of only the vertex corrections,  here $f_\q(\omega)$ provides a measure of the the combined impact of both the electron self-energy and vertex corrections.
However, if we ignore the renormalization of the quasiparticle spectrum, i.e., $Z_\q \rightarrow 1$ and ${\bar \epsilon_\q} \rightarrow \epsilon_\q$, then
$f_\q(\omega)$ becomes identical to $f_\q^{(2c)}(\omega)$,  as expected.

Now, by taking into account the renormalization of the quasiparticle spectrum, the  phonon spectral function  can be expanded as
$B_\q(\omega)= B_\q^0(\omega) + B^{(a)}_\q(\omega) + B^{(b)}_\q(\omega)$ 
Here $B_\q^0$ represents the non-interacting case, as discussed above in Sec. (III), while $B^{(a)}_\q(\omega) = -\frac{1}{\pi} [\Im \Pi^{(a)}_\q(\omega )]  \frac{4\Omega_0^2}{(\omega^2-\Omega_0^2)^2}$ and  
$B^{(b)}_\q(\omega)=-\frac{1}{\pi} [\Im \Pi^{(b)}_\q(\omega )]  \frac{4\Omega_0^2}{(\omega^2-\Omega_0^2)^2}$ describe the low-energy $(|\omega| <\Omega_0)$ polaronic features by taking into account the impact of the renormalization of the quasiparticle spectrum  
 by excluding and including the contributions of the vertex corrections, respectively. Using Eqs. (\ref{Eq-pia}) and (\ref{Eq-pib}) the components of phonon spectral function representing the low-energy polaronic features can be expressed as,  

\bea
B_\q^{(a)}(\omega) = {\bar S^{(a)}_\q (\omega)} \left \{ \delta (\omega - {\bar \epsilon_\q} ) +   \delta (\omega + {\bar \epsilon_\q} )  \right \}   \:,
\label {Eq-Ba}
\eea

\bea
B_\q^{(b)}(\omega) =-{\bar S^{(a)}_\q (\omega)} \left \{f_\q(\omega) \delta (\omega - {\bar \epsilon_\q} ) +  f_{-\q}(-\omega)  \delta (\omega + {\bar \epsilon_\q} )  \right \}  \:,
\nonumber \\ 
\label {Eq-Bb}
\eea

\begin{figure}
\begin{center}
\hspace*{-2mm}
\resizebox{87mm}{87mm} {\psfig{figure=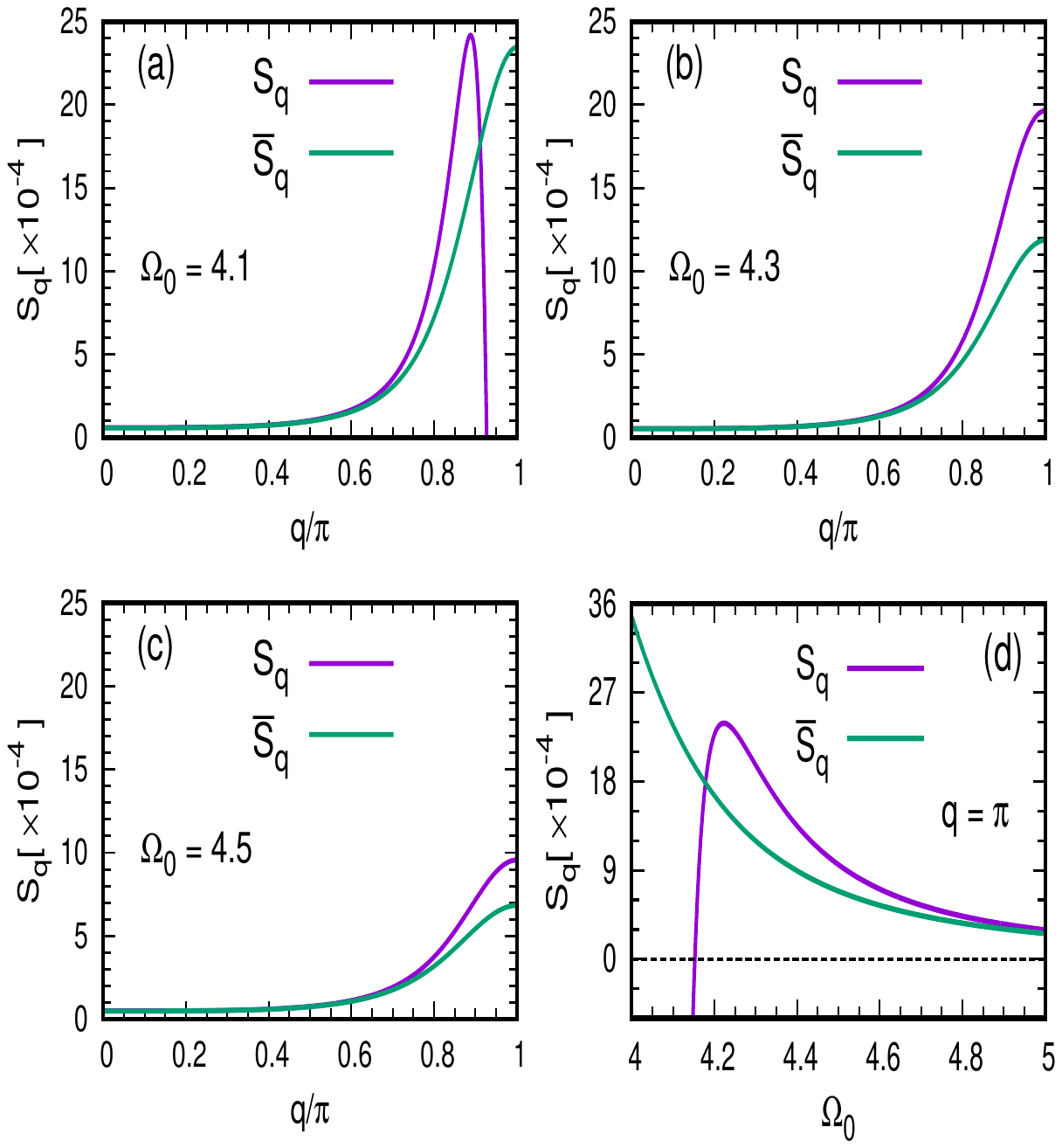}} 
\end{center}
\vspace*{-25mm}
\caption{ Comparison of the  phonon spectral weights  $S_\q$ and $\bar S_\q$ associated with the low-energy polaronic features, 
as obtained with the bare and the renormalized quasiparticle dispersions, respectively. 
While $S_\q \approx \bar S_\q$  for the small-$\q$ modes, irrespective of the value of $\Omega_0$,  on the other hand, 
the phonon spectral weight for the large-$\q$ modes near the zone boundary is strongly affected by the  renormalization of quasiparticle spectrum  both qualitatively and quantitatively as $\Omega_0 \rightarrow W$ in the antiadiabatic regime  (a, b, c) .  These include, e.g., 
resolving the issue of the unphysical negative spectral weight,  as demonstrated for $\Omega_0=4.1$ (a), and significant suppression in the spectral weight,  as demonstrated for  $\Omega_0=4.3$ (b) and  4.5 (c).  However, with increasing $\Omega_0$ the impact of the renormalization of the quasiparticle spectrum
for the large-$\q$ modes  weakens, as demonstrated for $\q=\pi$ mode (d) which eventually results in ${\bar S_\q} \approx  S_\q$ in  the entire Brillouin zone. These results have been obtained with parameters $g=0.5,  N=1000$.}
\label{Fig-q-Sq}
\end{figure}

\noindent where 
 ${\bar S^{(a)}_\q(\omega)}= \frac { g^2}{N}  Z_0 Z_\q \frac{4\Omega_0^2}{(\omega^2-\Omega_0^2)^2}$. 
Now, owing to  the $\omega-$symmetric nature of the phonon spectral function we focus only in region $\omega > 0$, where $B_\q^{(b)} (\omega)  = -f_\q(\omega) B_\q^{(a)} (\omega)$. Therefore, the net phonon spectral function representing the low-energy polaronic features can be written as  $B_\q (\omega) = B_\q^{(a)} (\omega) + B_\q^{(b)} (\omega) = B_\q^{(a)}(\omega) [1-f_\q(\omega) ]= {\bar S_\q}(\omega) \delta (\omega -{\bar \epsilon_\q})$, where ${\bar S_\q} (\omega) =  {\bar S_\q}^{(a)}(\omega)  [1-f_\q(\omega) ]$, so that ${\bar S_\q}= {\bar S_\q}(\omega={\bar \epsilon_\q})$ represents the net phonon spectral weight of the polaron by taking into account the renormalization of the quasiparticle spectrum.   Now, since $ {\bar  S_\q^{(a)}} (\omega={\bar \epsilon_\q}) =  \frac { g^2}{N}  Z_0 Z_\q \frac{4\Omega_0^2}{({\bar \epsilon_\q}^2-\Omega_0^2)^2}$ is always positive and finite as  $\bar \epsilon_\q < \Omega_0$, 
and $f_\q (\omega={\bar \epsilon_\q}) <0$ in the low-energy regime ($\omega < \Omega_0$), as obvious from Eq. ( \ref{Eq-Fq-renvertex}), therefore the net phonon spectral weight (${\bar S_\q}$) associated with the polaron always remains positive for all the $\q-$modes throughout the entire Brillouin zone, as demonstrated in Fig.  (\ref{Fig-q-Sq}) for different values of $\Omega_0$.  This is in sharp contrast to the unphysical negative phonon spectral weight ( $S_\q $ )  for the large-$\q$ modes near the zone boundary as $\Omega_0 \rightarrow W$
when the renormalization of the quasiparticle dispersion is ignored, as demonstrated in Fig. (\ref{Fig-q-Sq}) for $\Omega_0=4.1$.  
Furthermore,  a comparison of the results for $S_\q$ and $\bar S_\q$ also highlights the momentum-dependent impact of the renormalization of the  quasiparticle dispersion on the phonon spectral weight of polaron. As shown in Fig.  (\ref{Fig-q-Sq}), while $S_\q \approx  {\bar S_\q}$ for the small-$\q$ modes irrespective of the values of $\Omega_0$, 
on the other hand, the phonon spectral weight for the large-$\q$ modes is strongly affected by the renormalization of the quasiparticle spectrum when $\Omega_0 \rightarrow W$ in the antiadibatic regime. 
For example, apart from resolving the issue of the unphysical negative spectral weight, the renormalization of the quasiparticle spectrum also results in significant suppression of the phonon spectral weight for the large$-\q$ modes when $ \Omega_0 \rightarrow W$, as demonstrated for $\Omega_0= 4.3, 4.5$. 
However, with increasing $\Omega_0$ the impact of the renormalization for the large-$\q$ modes also weakens, as shown for the $\q=\pi$ modes in Fig. (\ref{Fig-q-Sq}). In other words, for large $\Omega_0$ we get  ${\bar S_\q} \approx S_\q$ in the entire Brillouin zone. In other words,  our weak-coupling perturbative analysis, where renormalization of the quasiparticle spectrum is ignored, as discussed  above in Sections II and III,  provides a reliable estimation of the impact of the many-body effects on the phonon spectral function of the polaron  in the antiadiabatic regime except when $\Omega_0 \rightarrow W$.  Our result for the weak impact of the renormalization on the quasiparticle spectrum in the  antiadiabatic regime ($\Omega_0 > W$) is also consistent with the  prediction of the exact diagonalization technique\cite{Fehske-PRB-1997}.

\section {conclusions} We have  carried out a weak-coupling analysis of  the impact of the many-body effects on the phonon spectral function of the Holstein polaron in one-dimension in the antiadiabatic regime. The many-body effects were incorporated  by taking into account the momentum and frequency dependent contributions of the leading-order electron self-energy and vertex corrections.  Our investigation has demonstrated the opposite impact of the  contributions from the electron self-energy and vertex corrections on the spectral weight associated with the polaronic features in the phonon spectral function.  While the polaronic spectral weight is suppressed due to contribution from the electron self-energy, on the other hand,  the same is enhanced due to  contribution from the vertex corrections. The two opposite  contributions  cancel each other exactly for the phonon modes  with zero wave vector  ($\q=0$)  as a consequence of the charge-conserving Ward identity which results in an identical vanishing of the net many-body correction.  With increasing $\q$, while strength of the  contributions from both the electron self-energy as well as the  vertex corrections increases,  the strength of the net many-body effects is governed by the interesting $\q$-dependent evolution of the two opposite contributions.  In the long-wavelength regime $(\q \rightarrow 0$) the two opposite contributions remain nearly equal and therefore almost cancel each other thereby resulting in a weak impact of the net many-body effects. As a result   
the polaronic spectral weight for the small-$\q$ modes  remains nearly unaltered by the  many-body effects.   On the other hand, for the large-$\q$ modes, specially near the zone boundary, we have found that the strength of the contribution from the electron self-energy increases rapidly in comparison to those from the vertex corrections  which result in  a  substantial suppression of the polaronic spectral weight in the phonon spectral function.  

We have also studied the impact of the  electron-phonon coupling driven renomalization of quasiparticle spectrum on the phonon spectral function of polaron. We have found  that our weak coupling perturbative analysis, where the renormalization of the quasiparticle spectrum is ignored,  provides a good estimation of the impact of the many-body  effects on the phonon spectral function of the polaron in the antiadiabatic regime when the  phonon energy ($\Omega_0$) remains sufficiently larger than the bandwidth ($W$) of the bare electronic dispersion.  However, in the limiting cases when  $\Omega_0 \rightarrow W$, we have found that  renomalization of the quasiparticle spectrum must be taken into account for an accurate estimation of the impact of many-body effects for the large$-\q$ modes. It would be interesting to study how the impact of many-body effects will be affected by taking into account the realistic features of the physical systems such as finite charge carrier density, dispersive phonon spectrum, momentum-dependent electron-phonon coupling, etc., which requires to extend our investigations using some more complicated models. This constitutes a non-trivial extension of our work in the near future.

\end{document}